\documentclass[twocolumn,showpacs,preprintnumbers,amsmath,amssymb]{revtex4}

\usepackage[dvips]{color}
\usepackage{graphicx}

\begin{document}

\title{Local quantum criticality in confined fermions on optical lattices}

\author{M. Rigol}
\affiliation{Institut f\"ur Theoretische Physik III, Universit\"at Stuttgart, 
Pfaffenwaldring 57, D-70550 Stuttgart, Germany.}
\author{A. Muramatsu}
\affiliation{Institut f\"ur Theoretische Physik III, Universit\"at Stuttgart, 
Pfaffenwaldring 57, D-70550 Stuttgart, Germany.}
\author{G.G. Batrouni}
\affiliation{Institut Non-Lin\'eaire de Nice, Universit\'e de Nice--Sophia
Antipolis, 1361 route des Lucioles, 06560 Valbonne, France}
\author{R.T. Scalettar}
\affiliation{Physics Department, University of California, Davis, CA 95616}

\begin{abstract}
Using quantum Monte Carlo simulations, we show that the
one-dimensional fermionic Hubbard model in a harmonic potential
displays quantum critical behavior at the boundaries of a
Mott-insulating region. A local compressibility defined to
characterize the Mott-insulating phase has a non-trivial critical
exponent. Both the local compressibility and the variance of the local
density show universality with respect to the confining potential. We
determine a generic phase diagram, that allows the prediction of the phases
to be observed in experiments with ultracold fermionic atoms trapped
on optical lattices.
\end{abstract}
\maketitle

The Mott metal-insulator transition (MMIT), a paradigm of strong
correlations, was recently realized in ultracold atoms confined on an
optical lattice \cite{greiner02}. Due to the fact that the atoms
interact only via a contact potential, this system constitutes the
most direct experimental realization of the Hubbard model which is the
prototype generally used to study the MMIT. Whereas optical lattices 
contain bosonic atoms, recent progress in cooling
techniques allow fermionic systems to go well below the degeneracy
temperature \cite{hadzibabic02,roati02}, such that even superfluidity
appears within reach \cite{ohara02}. It is therefore, to be expected
that soon a fermionic MMIT will be realized on an optical lattice,
offering the possibility to confront in a controlled way our knowledge
of the MMIT in solid-state systems, without extrinsic effects always
present there. This possibility is especially important since the MMIT 
is not only a long standing problem in condensed matter physics, but has 
also received renewed attention in recent
years due to its different manifestations in a number of transition
metal oxides, the most prominent being high temperature
superconductors \cite{imada98}.

Motivated by the possibility of such crossfertilization, we performed
QMC simulations for the ground-state of a one-dimensional Hubbard
model with a harmonic potential, as in experiments with ultracold
atoms, confining spin 1/2 fermions. The one-dimensional case was
chosen since in one dimension, the quantum critical properties for the
unconfined system are well characterized by the {\em Bethe Ansatz}
solution where, in particular, the global compressibility $\kappa \sim
\partial n/\partial \mu$ diverges as $\delta^{-1}$, where $\delta = 1
- n$, and $n$ is the expectation value of the density
\cite{usuki89}. However, as shown theoretically \cite{jaksch98} and
numerically \cite{batrouni02}, in the presence of
a confining potential the Mott-insulating phase is 
restricted to a domain that coexists with a compressible phase. This
is in contrast to the global character typical of solid state systems. 
We show in this Letter, that a properly defined
local compressibility displays critical behavior on approaching the
edges of the Mott-insulating phase, revealing a new critical
exponent. Furthermore, it is shown that both the variance of the local
density, $\Delta_i \equiv <n_i^2> - <n_i>^2$, 
and the local
compressibility as functions of the local density $n_i$, are
independent of the confining potential for $n_i \rightarrow 1$. The
exponents are also universal with respect to the strength of the
interaction.

The Hamiltonian studied is as follows:
\begin{eqnarray}
\label{Hamiltonian}
H & = & -t \sum_{i,\sigma} 
\left( c^\dagger_{i\sigma} c^{}_{i+1 \sigma} + h.c. \right)
+ U \sum_i n_{i \uparrow} n_{i \downarrow} 
\nonumber \\ & &
+ \left( \frac{2}{N} \right)^\alpha V_\alpha \sum_{i \sigma} 
\left(i - \frac{N}{2} \right)^\alpha n_{i \sigma},
\end{eqnarray}
where $c^\dagger_{i\sigma}$ and $c^{}_{i\sigma}$ are creation and
annihilation operators, respectively, for a fermion on site $i$ with
spin $\sigma = \uparrow, \downarrow$. The local density per spin is
$n_{i \sigma} = c^\dagger_{i\sigma} c^{}_{i \sigma}$.  The contact
interaction is repulsive ($U>0$) and the last term models the
potential of the magneto-optic trap. The QMC simulations were performed
using a projector algorithm \cite{sugiyama86,sorella89,loh92,imada93}, 
which applies $\exp (-\theta H)$ to a trial wavefunction (in our case 
the solution for $U=0$). A projector parameter $\theta \simeq 20/t$ suffices to 
reach well converged values of the observables discussed here. A time slice of
$\Delta \tau = 0.05/t$ was used in general.

\begin{figure}[h]
\includegraphics[width=.43\textwidth]{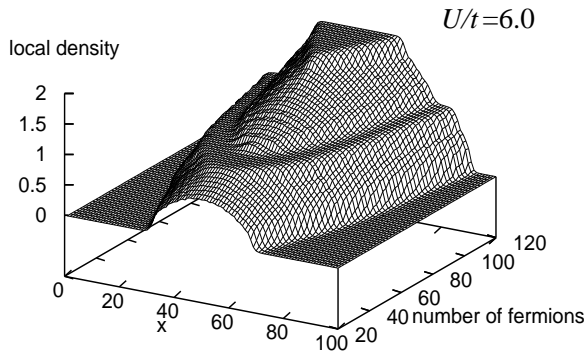}
\caption[]{Density profiles along the trap for different fillings.
Flat terraces are the Mott insulating regions.}
\label{dens3D}
\end{figure}

Figure \ref{dens3D} shows density profiles along a harmonic trap ($\alpha=2$) for
different fillings such that the system goes from an entirely metallic
phase to a phase with insulating regions due to full occupancy of the
sites, coexisting with metallic regions. At this point, such an 
identification is only based on the occupation number, using our knowledge 
from unconfined periodic systems. 
Although it is natural to identify the regions 
with $n=1$ as insulating phases, the global compressibility used in unconfined systems
is not a useful order parameter to characterize the phases
due to the coexistence of local compressible regions with incompressible ones.
A first quantity that can be used instead is
the variance of the site density $\Delta_i$, since on entering the
Mott-insulating region a suppression of double occupancy should occur,
leading to a decrease of the variance.

\begin{figure}[h]
\includegraphics[width=.42\textwidth]{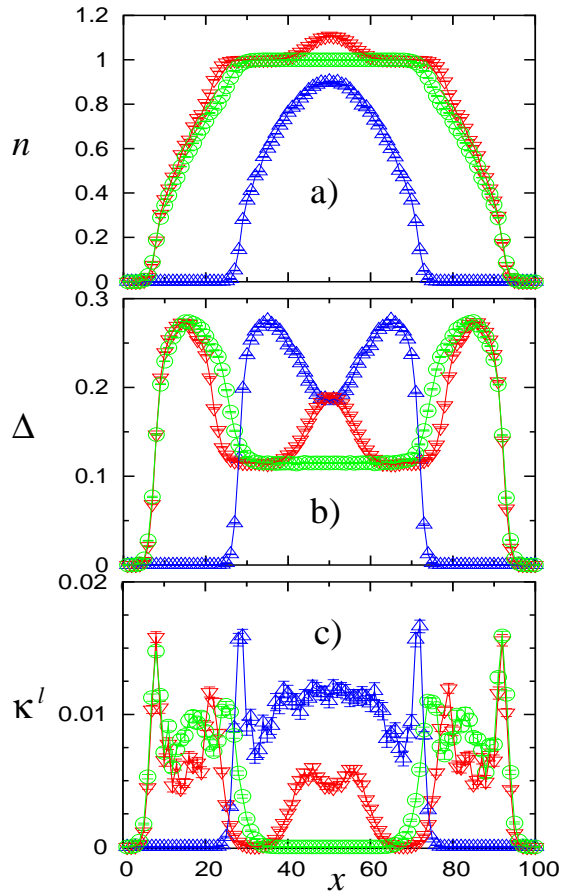}
\caption[]{(a) Density profiles for (\textcolor{blue}{$\triangle$}) 
$N_f = 30$, $U=6t$ and $V_2=15t$,
(\textcolor{green}{$\bigcirc$}) $N_f = 70$, $U=6t$ and
$V_2=6.25t$, and (\textcolor{red}{$\bigtriangledown$}) $N_f = 74$, 
$U=6t$ and $V_2=7t$. (b) Variance of the local density
(c) Local compressibility $\kappa^l$ as defined in eq.\
(\ref{localc}).}
\label{dens-var-comp}
\end{figure}

Figure \ref{dens-var-comp} (a) shows three characteristic density profiles.
In all of them, $U/t=6$ but depending on the
filling ($N_f$) and strength of the potential $V_2$, we obtain (i) an
approximately parabolic density profile, indicating that the whole
system is in a metallic phase.
(ii) Increasing the number of particles, a Mott-plateau develops in
the center of the system, and finally, (iii) with still a higher
filling, a new metallic phase develops in the center of the
plateau. The potential $V_2$ was varied in order to obtain well
developed phases. The corresponding profiles for the variance $\Delta$
are shown in Fig.\ref{dens-var-comp}(b). In general, a suppression of
$\Delta$ is present in the regions where the density profile shows
a plateau. However, although in the region with $n_i = 1$ the variance 
is lower than in the regions surrounding it, it does not
vanish. Therefore, while the decrease of $\Delta$ is a signature for
the Mott-insulating phase, still a clearer distinction is needed.  For
this purpose we introduce a {\em local} compressibility defined as
follows:
\begin{equation}
\label{localc}
\kappa_i^l = \sum_{\mid j \mid \leq \, l (U)} \chi_{i,i+j} \; ,
\end{equation}
where $\chi_{i,j}$ is the density-density correlation function. We
take the length scale $l(U) \simeq b \, \xi (U)$, where $\xi (U)$ is
the correlation length given by $\chi_{i,j}$ in the {\em unconfined}
system at half filling for the given value of $U$. The value of $b$ is
such that $\kappa^l$ becomes insensitive to the value chosen.  In
general we have $b \sim 5 - 10$, with $\xi (U) \sim a$ ($a$ is the
lattice constant) for the values of $U$ used here.  The local
compressibility thus gives the response to a constant shift of the
potential over a finite range but over distances larger than  $\xi (U)$ 
in the periodic, unconfined system. If a region is
in a Mott-insulating phase, and hence incompressible, 
no density response over distances larger than $\xi$ is expected, leading 
to $\kappa_i^l=0$. Figure \ref{dens-var-comp}(c) shows the
profile along the trap of $\kappa_i^l$, where the compressibility
becomes zero in the outlying regions, where no particles are present
and also where a Mott plateau is present. Therefore, the local
compressibility defined here serves as a genuine local order parameter
to describe the insulating regions that coexist, in general, with a
surrounding metallic zone or even with metallic intrusions, 
beyond the intuitive pictures on the basis of the density profiles.  

Now that phases can be characterized quantitatively, we concentrate on
the regions where the system goes from one phase to another. 
Criticality can arise, despite the microscopic spatial size,
due to the extension in
imaginary time that reaches a thermodynamic limit at $T=0$, very much
like the case of the single impurity Kondo problem \cite{yuval70},
where long-range interactions in imaginary time appear for the local
degree of freedom as a result of the interaction with the rest of the system.
Recent experiments leading to a MMIT \cite{greiner02} consider a system 
with linear dimension $\sim 65 a$, i.e. still in this microscopic range. 
An intriguing future question, for both theory
and experiment, will be the role of spatial dimension
in the critical behavior of systems in the thermodynamic limit.

\begin{figure}[h]
\includegraphics[width=.42\textwidth]{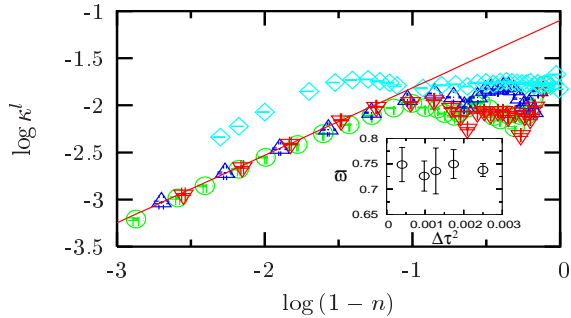}
\caption[]{The local compressibility $\kappa^l$ {\em vs.} 
$\delta=1-n$ at $\delta \rightarrow 0$ for
(\textcolor{blue}{$\triangle$}) $N_f = 70$, $U=8t$ and
$V_2=6.25t$; (\textcolor{red}{$\bigtriangledown$}) $N_f = 70$, $U=6t$
and $V_2=6.25t$; (\textcolor{green}{$\bigcirc$}) $N_f=72$, $U=6 t$
and a quartic potential with $V_4=6.25t$; (\textcolor{cyan}{$\Diamond$})
unconfined periodic system with $U=6t$.
Inset: Dependence of the
critical exponent $\varpi$ on $\Delta \tau^2$.}
\label{kappavsn}
\end{figure}

Figure \ref{kappavsn} shows the local compressibility {\em vs.} 
$\delta$ for $\delta \rightarrow 0$ in a double logarithmic plot. A
power law $\kappa^l \sim \delta^\varpi$ is obtained, with $\varpi <
1$, such that a divergence results in its derivative with respect to
$n$, showing that critical fluctuations are present in this region.
Since the QMC simulation is affected by systematic errors due to
discretization in imaginary time, it is important to consider the
limit $\Delta \tau \rightarrow 0$ in determining the critical
exponent. The inset in Fig. \ref{kappavsn} shows such an
extrapolation leading to $\varpi \simeq 0.68 - 0.78$.  At this point
we should remark that the presence of the harmonic potential allows the 
determination of the density dependence of various quantities with 
unprecedented
detail on feasible system sizes as opposed to unconfined periodic
systems, where systems with $10^3 - 10^4$ sites would be necessary to
allow for similar variations in density. In addition to the power law
behavior, Fig. \ref{kappavsn} shows that for $\delta
\rightarrow 0$, the local compressibility of systems with a harmonic
potential but different strengths of the interaction or even with a
quartic confining potential, collapse on the same curve. 
Hence, universal behavior as expected for critical
phenomena is observed also in this case.  This fact is particularly
important with regard to experiments, since it implies that the
observation of criticality should be possible for realistic confining
potentials, and not only restricted to perfect harmonic ones, as usually
used in theoretical calculations. 
However, Fig. \ref{kappavsn} shows also that the unconfined case
departs from all the others. Up to the largest systems we simulated
(400 sites), we observe an increasing slope rather than the power law of the
confined systems. 

\begin{figure}[h]
\includegraphics[width=.44\textwidth]{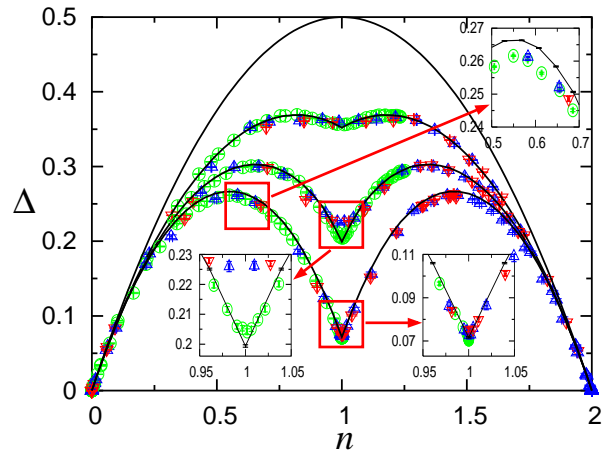}
\caption[]{a) Variance $\Delta$ {\em vs.} $n$ for
(\textcolor{green}{$\bigcirc$}) harmonic potential $V_2=6.25 t$ with $N=100$; 
(\textcolor{blue}{$\triangle$}) quartic
potential $V_4=15.82 t$ with $N=150$;
(\textcolor{red}{$\bigtriangledown$}) harmonic potential $V_2=10 t$ + 
cubic $V_3 = 2.5 t$ + quartic $V_4 = 7.5 t$ with $N=50$; and (full line) 
unconfined periodic
potential with $N=102$ sites. The curves correspond from top to bottom to $U/t = 0, 2, 4,
8$. For a discussion of the insets, see text.}
\label{DeltaVsnAll}
\end{figure}

Having shown that the local compressibility displays universality on
approaching a Mott-insulating region, we consider the variance
$\Delta$ as a function of the density $n$ for various values of $U$
and different confining potentials. Figure \ref{DeltaVsnAll} shows
$\Delta$ {\em vs}. $n$ for a variety of systems, where not only the
number of particles and the size of the system are changed, but also
different forms of the confining potential were used.  Here we
considered a harmonic potential, a quartic one, and a superposition of
a harmonic, a cubic and a quartic one, such that even reflection
symmetry across the center of the system is broken. It appears at
first glance that the data can only be distinguished by the strength
of the interaction $U$, showing that the variance is rather
insensitive to the form of the potential.  The different insets,
however, show that a close examination leads to the conclusion that
only near $n=1$ and only in the situations where at $n=1$ a
Mott-insulator exists, universality sets in.
The inset for $n$ around 0.6 and $U=8t$, shows that the unconfined
system has different variance from the others albeit very close on a
raw scale. This difference is well beyond the error bars. Also the inset 
around $n=1$ and for $U=4t$,
shows that systems that do not form a Mott insulating phase in spite of
reaching a density $n=1$, have a different variance from those having
a Mott-insulator. The features above show that even a
very local quantity like the variance cannot be accurately described using
a local density approximation (Thomas-Fermi) \cite{butts97}, and can lead to 
even qualitatively wrong results, as for $U=4t$ and $n=1$, where such an 
approximation would predict a Mott-insulator instead of a metal as in our 
simulations.  
Only the case where all systems have a
Mott-insulating phase at $n=1$ ($U=8t$), shows universal behavior
independent of the potential, a universality that encompasses also the 
unconfined systems. For the unconfined system, the behavior of the 
variance can be examined with Bethe-{\em Ansatz} \cite{lieb68} in the limit 
$\delta \rightarrow 0$. In this
limit and to leading order in $\delta$, the ground state energy is
given by \cite{schadschneider91} $E_0 (\delta)/N - E_0 (\delta=0)/N
\propto \delta$, such that the double occupancy, which can be obtained
as the derivative of the ground-state energy with respect to $U$,
will also converge as $\delta$ towards its value at half-filling. Such
behavior is also obtained in our case as shown by the inset at $n=1$
($U=8t$) in Fig. \ref{DeltaVsnAll}. Detailed data for the variance
close to $n=1$ will be presented elsewhere.

\begin{figure}[h]
\includegraphics[width=.30\textwidth,height=.22\textwidth]{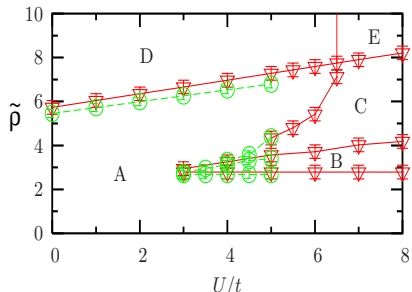}
\vspace{-0.5cm}\caption[]{Phase diagram for a system with $N=100$ 
(\textcolor{red}{$\bigtriangledown$}) and $N = 150$ 
(\textcolor{green}{$\bigcirc$}) sites. The phases are explained 
in the text.}
\label{PhaseD}
\end{figure}

Finally we consider the phase diagram of the system. As shown in Fig. 
\ref{dens-var-comp}, unlike the exponents,
phase boundaries seem to be rather sensitive to the choice of 
potential, number of particles and strength of the interaction. As in
the unconfined case, we would expect to be able to relate systems with
different number of particles and/or sizes by their density. Given the
harmonic potential, a characteristic length (in units of the lattice
constant) is given by $N \left(4 V_2/t \right)^{-1/2}$, such that a
characteristic density can be defined. Figure \ref{PhaseD} shows that
the characteristic density $\tilde{\rho}=\rho \sqrt{4 V_2/t}$ 
($\rho=N_f/N$) is a meaningful 
quantity to characterize the phase diagram. 
There, the phase diagrams for two
systems with different sizes ($N = 100$ and $N = 150$) and different
strength of the harmonic potential ($V_2=15 t$ and $V_2=11.25 t$
respectively) are depicted showing that such a scaling allows to
compare systems with different sizes, different number of particles,
and different strength of the potential. This makes it possible to relate
the results of numerical simulations to much larger experimental
systems. The different phases obtained are: A pure metal without
insulating regions (A), a Mott-insulator at the center of the trap
(B), a metallic intrusion at the center of a Mott-insulator (C), a
``band insulator'' (i.e. with $n=2$) at the center of the trap
surrounded by a metal (D), and finally a ``band insulator'' surrounded
by a metal, surrounded by a Mott-insulator with the outermost region
being again a metal (E). Two features are remarkable here. The first 
one is that on varying the filling of the trap, a reentrant behavior is
observed for the phase A. The density profile shows a shoulder as can
be seen in Fig. \ref{dens3D} before reaching the plateau with $n=2$
but, as shown by the inset of Fig. \ref{DeltaVsnAll} for $U=4$ around 
$n=1$, it is possible to go through
a region with $n=1$ without reaching the value of the variance that
corresponds to a Mott-insulator.
The second intriguing feature is that the boundary between the regions A and B
remains at the same value of the characteristic density for all values
of $U$ that could be simulated.

In summary, on the basis of QMC simulations of the Hubbard model with
a harmonic potential, we found a number of new and unexpected features
for the MMIT. ({\it i}) A local compressibility
$\kappa^l$ that appropriately characterizes Mott-insulating regions, shows
critical behavior on entering those regions. 
Due to the microscopic nature
of the phases, spatial correlations appear not to contribute to the
critical behavior discussed here. 
This is a new form of MMIT, not observed so far in simple periodic systems,
that might be realized in fermionic gases trapped on optical lattices.
Therefore, our observation adds a new aspect to this long-standing problem in
condensed matter physics. We expect that a similar local
critical behavior will arise in higher dimensions as long as the spatial
extent of the Mott domain remains finite. 
({\it ii}) Universal behavior is found for
$\kappa^l$ for $n \rightarrow 1$, independent of the confining
potential and/or strength of the interaction, excluding, however, the 
unconfined case. 
Also universal behavior is
found for the variance $\Delta$ when $n \rightarrow 1$. In this case, 
this behavior is shared by the 
unconfined model. ({\it iii}) Finally, a proper scaling form
for a characteristic density is introduced that leads to a generic
phase diagram, with interesting features, as described in the
paragraph above.

We wish to thank HLR-Stuttgart (Project DynMet) for allocation of
computer time.  We gratefully acknowledge financial support from the
LFSP Nanomaterialien, a PROCOPE grant (France-Germany), NSF-CNRS-12929,
NFS-DMR-9985978, NSF-DMR-0312261, and NFS-INT-0124863.  
We are indebted to M. Arikawa, F. G\"ohmann, and
A. Schadschneider for instructive discussions on Bethe-Ansatz, and to
M. Feldbacher, B. O. Cult and F.F. Assaad for interesting
discussions.

\end{document}